# Sublime Transfer Printing of Three-Dimensional Nanostructure Ensembles

Lei Chen[1,2], Hao Wang[4,5], Wang Zhang[1,2,3*], Fu Fan[3], Peng Liu[6], Xiaoxue Bi[3], John You En Chan[1], Cheng-Feng Pan[1], Bochang Wu[1], Zhengchao Liu[1], Rou Yun Teo[1,2], Hongtao Wang[1], Huigao Duan[3*], Joel K.W. Yang[1,2*]

[1]Engineering Product Development, Singapore University of Technology and Design, Singapore 487372, Singapore. [2]Singapore-HUJ Alliance for Research and Enterprise (SHARE), The Smart Grippers for Soft Robotics (SGSR) Programme, Campus for Research Excellence and Technological Enterprise (CREATE), Singapore 138602, Singapore. [3]College of Mechanical and Vehicle Engineering, Hunan University, Changsha 410082, P. R. China. [4]School of Instrumentation and Optoelectronic Engineering, Beihang University, Beijing 100191, P. R. China. [5]Hangzhou International Innovation Institute, Beihang University, Hangzhou 311115, P. R. China. [6]School of Mechanical Engineering, Hunan University of Science and Technology, Xiangtan 411201, P. R. China.

L. C. and Hao Wang contributed equally to this work.

*Corresponding authors: zhangwang07@hnu.edu.cn (Z. W.); duanhg@hnu.edu.cn (H. D.); joel_yang@sutd.edu.sg (J.K.W.Y.)

***Abstract***

High-resolution three-dimensional (3D) nanostructures for visible-light photon manipulation provide unique and bespoke capabilities in optics and photonics. However subwavelength nanofabrication and reliable ensemble manipulation of the 3D prints onto arbitrary substrates remain challenging. Here, we introduce “sublime transfer” strategy tailored for transfer printing ensembles of delicate 3D printed nanostructures. This strategy enables conformal, damage-free integration of arrays of 3D structures on diverse substrates. Naphthalene acts as a transient stamp to encapsulate the structures during transfer and placement. We rely on the low sublimation temperature of naphthalene to release the structures reliably with nearly zero stress, preventing mechanical damage and positional misalignment. This approach is broadly applicable to integrate diverse nanostructures and photonic devices onto various substrates, and enabling inorganic architectures through ensemble uniform post-processing, including 2.5D photonic crystals on flexible PDMS, diffractive optical elements on curved lenses, spiral phase plates on CMOS chips, multilayer achromatic metalens on optical fiber facet, as well as 3D glass photonic crystals and optical topological resonators on anti-stiction quartz.

## Introduction

Realization of high-resolution 3D nanostructures in the visible spectrum for photon manipulation has become a key frontier in optics and photonics[1,2]. Compared with 2D or quasi-2D structures, 3D nanostructures provide higher structural degrees of freedom at subwavelength scales, enabling multi-parameter manipulation over photon propagation, polarization and local electromagnetic fields[2,3]. These structured media allow tailoring of light-matter interactions, revisiting fundamental optical laws, and exploring novel device concepts, such as 3D photonic crystals[1,4,5], metamaterials[6], chiral helices[7,8], plasmonic nanostructures[9], and all-dielectric topological insulators[10]. Two-photon polymerization lithography (TPL) achieves sub-diffraction-limited fabrication by nonlinear two-photon absorption and voxel-by-voxel overlapping assembly, making it a representative method for these complex 3D nanostructures[11-15]. However, the resolution achievable by direct TPL is limited by optical diffraction, material mechanics, and proximity effects, restricting its ability to construct subwavelength features with high precision[16,17]. Its reliance on planar rigid substrates also limits 3D fabrication on flexible or complex surfaces, constraining integration with heterogeneous and freeform photonic devices[18].

To push the resolution and functionality of TPL-fabricated 3D photonic nanostructures beyond the capabilities of the lithographic system/process, recent efforts have sought to incorporate “print-and-shrink” post-processing strategies such as thermal shrinkage or pyrolysis[19-25]. Substantial advancements have been achieved in high-resolution polymeric architectures, as well as functional 3D nanostructures such as oxides or metals[1,4,5,19,26-28]. However, nearly all existing concepts and approaches remain challenged by both spatially non-uniform shrinkage, leading to anisotropic deformation and structural distortion, and the limited compatibility with substrate types, which together hinder the realization of uniform, high-resolution 3D architectures across diverse platforms[20]. Isotropic shrinkage emerged as a promising strategy for achieving nanoscale uniform features beyond the resolution limits of 3D printing. However, current approaches are largely restricted to isolated, large-scale 3D structures, whereas extending shrinkage to ordered ensembles of 3D architectures remains challenging[16]. Such an extension requires preservation of spatial registration of 3D ensembles during transfer and post-processing for array-level fabrication and device integration. Transfer printing provides a potential route for manipulating ensembles of 3D structures and coupling them with the uniform shrinkage process, enabling integration on diverse substrates and on-chip photonic devices[29-32]. However, existing transfer printing strategies that use solid or liquid media as carriers to manipulate micropatterns, are limited to 2D structures and incompatible with complex or fragile 3D architectures and post-processing[18,29,33-36]. Thus far, achieving scalable 3D nanofabrication and manipulation of high-resolution structure ensembles remains a long-standing challenge in 3D photonics and micro-/nanomanufacturing.

To address these challenges, we introduce a “sublime transfer” strategy tailored for the manipulation and construction of 3D nanostructure ensembles onto substrates with flat and curved surfaces. This strategy enables the conformal transfer of ensembles of arbitrary 3D structures onto diverse substrates without damaging either the structures or the target surfaces, in contrast to previously reported transfer methods that induce stress, deformation, or positional instability (Supplementary Table 1). The key lies in the use of naphthalene as the carrier to completely encapsulate 3D structures during the pickup process and thereby preventing mechanical damage and positional shifts. Upon sublimation, naphthalene transitions directly into the gas phase, thus reliably releasing the 3D structures with negligible stress to the intricate structures, while eliminating solvent exposure and capillary forces. Therefore, placement onto a wider range of arbitrary substrates is now possible, including flexible, suspended, and non-stick surfaces. We performed theoretical analysis, mechanical tests and molecular dynamics simulation to further investigate the damage-free mechanism of this approach. We show that the method enables uniform print-and-shrink post-processing for the fabrication of high-resolution, uniform, and arbitrary inorganic 3D architectures. To show versatility, we transfer printed various nanostructures and photonic devices onto a variety of substrates to achieve structural colors on flexible substrates, diffractive optical elements (DOEs) on a curved surface, spiral phase plates on a microchip, multilayer achromatic metalens on a fiber facet, and defect-free, perfectly ordered 3D photonic crystals, and optical topological resonators on anti-stiction quartz, highlighting the potential impact of this method to optical applications and broader micro- and nanoscale systems.

## Results

### Concept of the sublime transfer strategy

Fig. 1a-b shows the concept and process flow of the sublime transfer strategy. 3D nanostructures are first printed at their designated positions using TPL on a donor substrate coated with an anti-stiction layer. Molten naphthalene at 120 °C is poured onto the substrate to encapsulate the ensemble/array of 3D structures. In the molten state, naphthalene flows into gaps, overhang and holes in the nanostructures to fully encapsulate without dislodging them from the substrate. Once solidified, the solid naphthalene sheet is sheared off the substrate with a light lateral force, carrying with it the full ensemble/array of structures in their original positions. This droplet can then be transferred by placing it onto any receiving substrate. Alignment is performed by translating the solid droplet under an inverted microscope to register the structures with the target substrate. The sample is kept at an elevated temperature (e.g. 50 °C) to accelerate sublimation, leaving behind only the 3D structures (Supplementary Fig. 1-2).

Our process minimizes stress concentration by encapsulating the 3D structures with molten naphthalene. Upon crystallization, the naphthalene bulk prevents damage and positional displacement of the 3D architectures during transfer. The naphthalene then transiently conforms to the target substrate through van der Waals (vdW) interactions before dissipating via direct sublimation into the gas phase. The sublimation process avoids liquid–vapor interface formation, eliminating capillary forces that may induce structural deformation or collapse, making it suited for high-aspect-ratio nanostructures. Crucially, the entire transfer process requires no chemical pretreatment of the receiver substrate. The process is residue-free and involves no harsh organic solvents (Supplementary Fig. 3). Notably, the naphthalene enables transfer onto low-surface-energy substrates, which is challenging for conventional switchable-adhesion stamps due to limited adhesion reduction during release. To show the strategy-enabled ensemble uniform shrinkage process, the transferred structures are released after naphthalene sublimation and heated above 450 °C to form high-resolution 3D structures, e.g. in inorganic materials (Fig. 1c, Supplementary Fig. 4 and Supplementary Table 2)[1,16].

The damage-free mechanism of the process was elucidated through theoretical analysis, adhesion tests and molecular dynamics simulation, as shown in Supplementary Discussion 1. These results indicate that near-zero interfacial adhesion between the low-energy donor substrate and 3D structures or naphthalene is critical for damage-free pick-up. Subsequent naphthalene sublimation reduces its interfacial adhesion to both structures and substrates to zero, enabling transfer to any surface. Notably, the uniform print-and-shrink process further relies on sufficiently weak interfacial interactions and adhesion between the 3D structures and the receiver substrate (Supplementary Fig. 5-9).

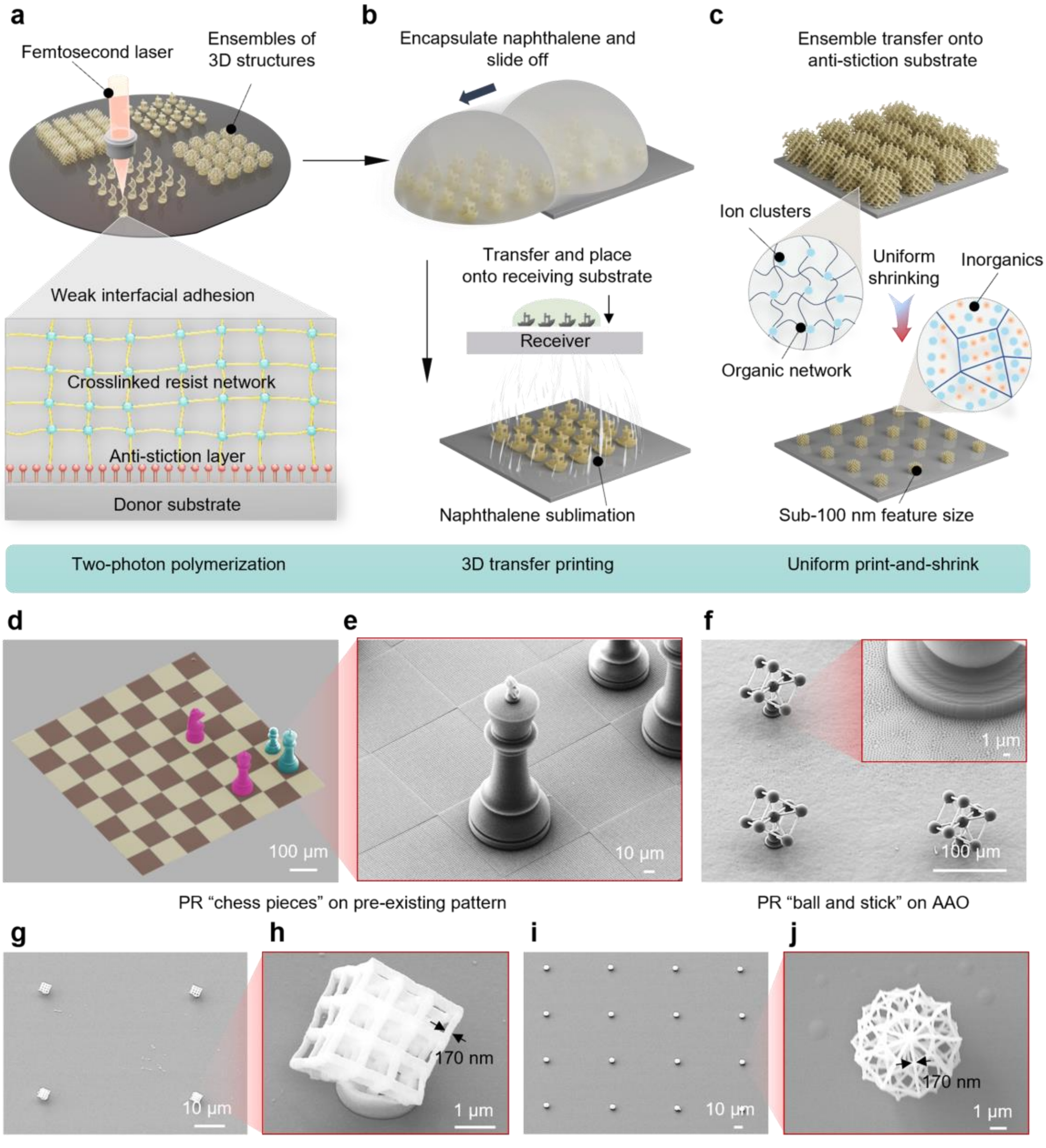

**Fig. 1. Ensemble nanofabrication and manipulation of 3D nanostructures on a variety of substrates via sublime transfer printing, aligned placement, and uniform print-and-shrink processes.** (a-b) Schematic of the two-photon polymerization lithography (TPL) and sublime transfer process. An ensemble of 3D structures fabricated on a low-surface-energy donor substrate with anti-stiction layer by TPL. 3D structures then encapsulated in molten naphthalene. After naphthalene crystallization, ensemble of 3D structures picked up and transferred onto receiving substrate, then released through naphthalene sublimation. (c) Schematic of ensemble uniform print-and-shrink process: 3D structures transferred onto low-surface-energy receiver substrate and released, followed by thermal decomposition to realize uniform shrinkage. (d-f) SEM image of transferred 3D structures onto pre-patterned 2.5D structures and anodic aluminum oxide (AAO) substrate. Chess pieces were aligned during transfer to underlying chess board (consisting of grid structures) and visualized using false-color rendering. PR refers to IP-DIP2 resist. (g-j) SEM image of $SiO_2$ and $TiO_2$ 3D nanostructures achieved after heating structures on the receiving substrate above 550 °C using previously reported custom resins and processes[1,4].

**Broad compatibility for complex 3D nanostructure construction**

To demonstrate the capability for aligned placement during transfer, we constructed a chess "endgame" model showing also the multi-material, heterogeneous integration outcome of transfer printing. Here chess pieces that were 3D printed in IP-DIP2 resist were aligned and transfer-printed onto the chess-board, consisting of pre-existing 2.5D photonic grids of a different material (Glass-Nano resin[4]), as shown in Fig. 1d-e and Supplementary Fig. 10a-c. The damage-free capability of the transfer process was then demonstrated by transferring "ball and stick" structures 3D printed in IP-DIP2 resist as shown in Fig. 1f[37]. These structures are particularly vulnerable to external forces and adhesion-induced deformation during transfer due to the low stiffness of the "sticks" with a diameter of only ~3 μm and stress concentration at the junctions. Using our transfer printing strategy, the structures were transferred onto several different types of substrates, including the tip/edge of a curved hypodermic needle (Supplementary Fig. 10d), suspended anodic aluminum oxide (AAO) templates (Fig. 1f), planar glass (Supplementary Fig. 10e), and fabric (Supplementary Fig. 10f). Scanning electron microscope (SEM) images show the successful transfer of these 3D structures in the right orientation, intact morphology, and precise alignment onto residue-free surfaces. We further demonstrated the strategy-enabled print-and-shrink process, achieving uniform shrinkage of ensembles of delicate structures. The initial feature sizes of silicon dioxide ($SiO_2$) and titanium dioxide ($TiO_2$) lattice arrays were 800 nm and 910 nm, respectively. Both structures underwent uniform (isotropic) shrinkage, achieving a minimum feature size of ~170 nm, yielding high-resolution, inorganic architectures (Fig. 1g–j and Supplementary Fig. 10g–h).

The transfer process enables the conformal manufacturing of microstructures onto nonplanar surfaces. For example, a planar chainmail structure was transferred onto the tip of a thumbtack, where conformal contact with the curved surface was achieved as naphthalene melted, recrystallized and wrapped around the surfaces (Fig. 2a). Discrete microarrays were also conformally transferred onto the uneven surface of a coin (Fig. 2b). Furthermore, structures that are difficult to directly fabricate using TPL, such as vertically aligned concentric rings, were realized by first printing the concentric rings in a planar configuration and then orienting the naphthalene layer vertically during sublimation (Fig. 2c).

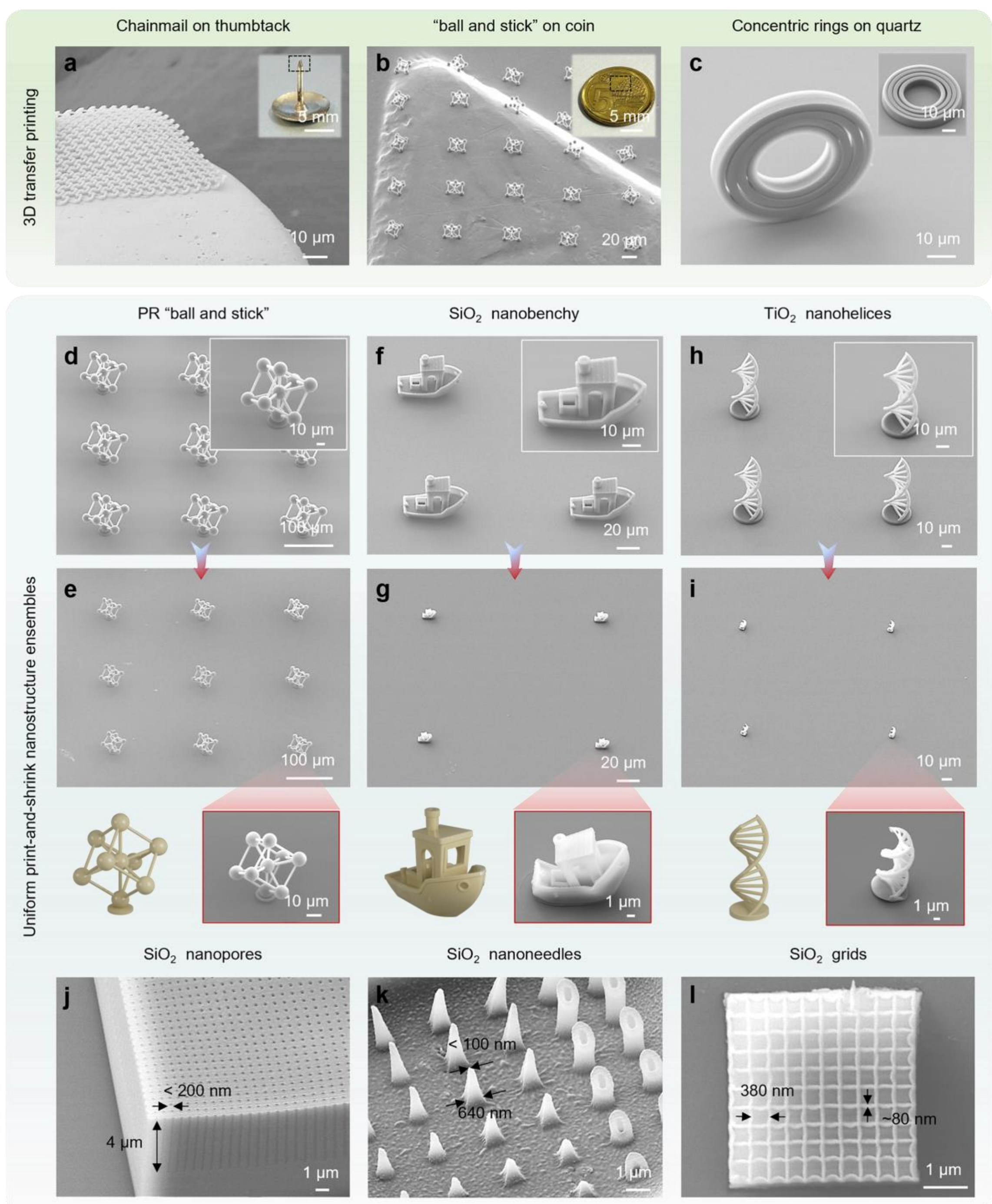

**Fig. 2. Demonstration of the transfer and uniform print-and-shrink process.** (a) SEM image of a planar chainmail structure transferred onto the tip of a thumbtack. (b) Discrete microarrays transferred onto a coin. (c) SEM image of vertically aligned concentric rings, the inset figure shows the original 3D-printed planar rings. (d-e) Original IP-DIP2 "ball and stick" arrays and after uniform print-and-shrink processing. (f-g) 2 × 2 arrays of $SiO_2$ nanobenchy print and shrink. (h-i) 2 × 2 arrays of $TiO_2$ nanohelices print and shrink. (j) $SiO_2$ nanopores. (k) $SiO_2$ nanoneedle arrays. Nanoneedles were printed with conical, four-sided pyramidal, and beveled geometries (left to right), with heights increasing from bottom to top. (l) 2.5D $SiO_2$ grid structures.

Next, we demonstrate the transfer printing-enabled strategy for achieving uniform shrinkage in detail via the shrinkage of structures fabricated from IP-DIP2 resist. The "ball and stick" arrays, after transfer onto anti-stiction substrates and thermal processing, exhibited uniform feature shrinkage by a factor of ~1.9 without array distortion (Fig. 2d–e). The shrinkage originates from mass loss during thermal decomposition, as confirmed by thermogravimetric analysis (Supplementary Fig. 11)[1,4]. Each element retains its central position during thermal processing in the array. This is verified using alignment markers fabricated simultaneously with the structures and transferred together, the shrinkage process in an array preserves the array pitch in physically isolated structures, or proportionally shrink both features and spacing in connected structures, with shrinkage occurring symmetrically about the geometric center (Supplementary Fig. 12). Such a geometrically defined center is no longer maintained in non-uniform shrinkage systems as directly print-and-shrink processing the 3D structures on low-surface-energy modified quartz substrates still results in non-uniform structures, reiterating the necessity of transfer printing (Supplementary Fig. 13)[16]. Moreover, the process enables the fabrication of various inorganic micro-/nanostructures with complex geometries, including 2 × 2 arrays of $SiO_2$ nanobenchy (Fig. 2f-g) and 2 × 2 arrays of $TiO_2$ nanohelices (Fig. 2h-i). High-aspect-ratio $SiO_2$ nanopores with sub-200 nm diameters and ~20:1 aspect ratios are achieved (Fig. 2j), together with high-resolution $SiO_2$ nanoneedle arrays with tip size of sub-100 nm (Fig. 2k), and 2.5D $SiO_2$ grid structures with feature size down to ~80 nm (Fig. 2l), In addition, complex functional 3D structures (e. g. Zinc oxide (ZnO) semiconductor architectures) are realized using different resists (Supplementary Fig. 14-15).

**Proof-of-principle photonic applications**

Photonic crystals devices with micro-nanostructures demonstrate promising applications in integrated optical components, anti-counterfeiting labels, and structural color printing[1]. However, integration across diverse substrates remains challenging. Our transfer process overcomes this limitation, enabling damage-free fabrication of complex photonic architectures on a variety of substrates. For example, 2.5D grid photonic structures can be conformally transferred onto stretchable and flexible substrates such as polydimethylsiloxane (PDMS). As shown in Fig. 3, we fabricated 2.5D grid structures with a base layer overlaid by submicron-scale grids, which act as color filters through light scattering and interference[38]. Colors are highly sensitive to grid geometry, including linewidth and height. By controlling laser power (30 mW), grid spacing (~2 μm), writing speed (10–20 mm/s in 1 μm/s steps), and number of layers (nominal height $h_2$ = 0.9–3.3 μm in 0.3 μm steps), we tuned the grids to fabricate an ensemble of color filters in the form of a color palette. Using this color palette as the base unit, we fabricated a large-area array (6.3 mm × 5.6 mm) consisting 6,930 color elements (Fig. 3a).

The palette was then transferred onto a PDMS substrate (Fig. 3b). Optical and SEM analyses confirmed that the micro-nanostructures of each filter preserved their morphology and relative positions within a positional accuracy of 1 μm (Fig. 3c-d and Supplementary Fig. 16). The spectra of representative colors (labeled *A*, *B*, and *C*) remaining unchanged, exhibiting only minor shifts, thereby demonstrating damage-free and reliable transfer printing (Fig. 3e). We further quantified the transfer yield of the 99 color filters in the color palette for filters ranging in size from 4 to 20 μm, achieving 100% yield (Fig. 3f). Based on this approach, we realized strain-insensitive structural colors that encode the "SUTD" letter patterns (Supplementary Fig. 17).

Our process also enables the integration of DOEs with curved lenses (e.g. contact lens), promising to combine the advantages of diffractive and freeform/curved-surface optics, enabling highly compact 3D optical systems and expanding the functional scope of conventional planar DOEs[39]. A DOE was designed and 3D-printed, then transferred it onto a convex glass lens (Fig. 3g and Supplementary Discussion 2). The DOE has a lateral size of 1 × 1 mm and comprises 1 million phase pixels, each with a feature size of 1 μm. When integrated onto the curved substrate, the DOE illuminated by a laser beam generates a holographic "SUTD" pattern projected into free space (Fig. 3h). Although curvature-dependent compensation algorithms were not applied in this proof-of-concept demonstration, the results are sufficient to indicate the potential of integrating planar DOEs with complex 3D curved optical surfaces.

To further demonstrate the heterogeneous integration capability of our process, spiral phase plates were fabricated on planar donor substrates and transfer-printed onto CMOS chips, devices that are capable of imparting helical phase profiles to the incident beam for vortex-beam generation with controlled orbital-angular-momentum states (Fig. 3i)[40]. Similarly, multilayer achromatic metalens arrays were transfer-printed onto plastic optical fibers, structures that possess high numerical aperture, broadband imaging potential at the fiber facet (Fig. 3j)[41]. Although these substrates could be written on directly, our approach avoids focal-plane drift during 3D printing, prevents solvent exposure, and offers a potential route for integrating oxide-based 3D structures onto functional devices in heterogeneous photonic packaging and on-chip photonics.

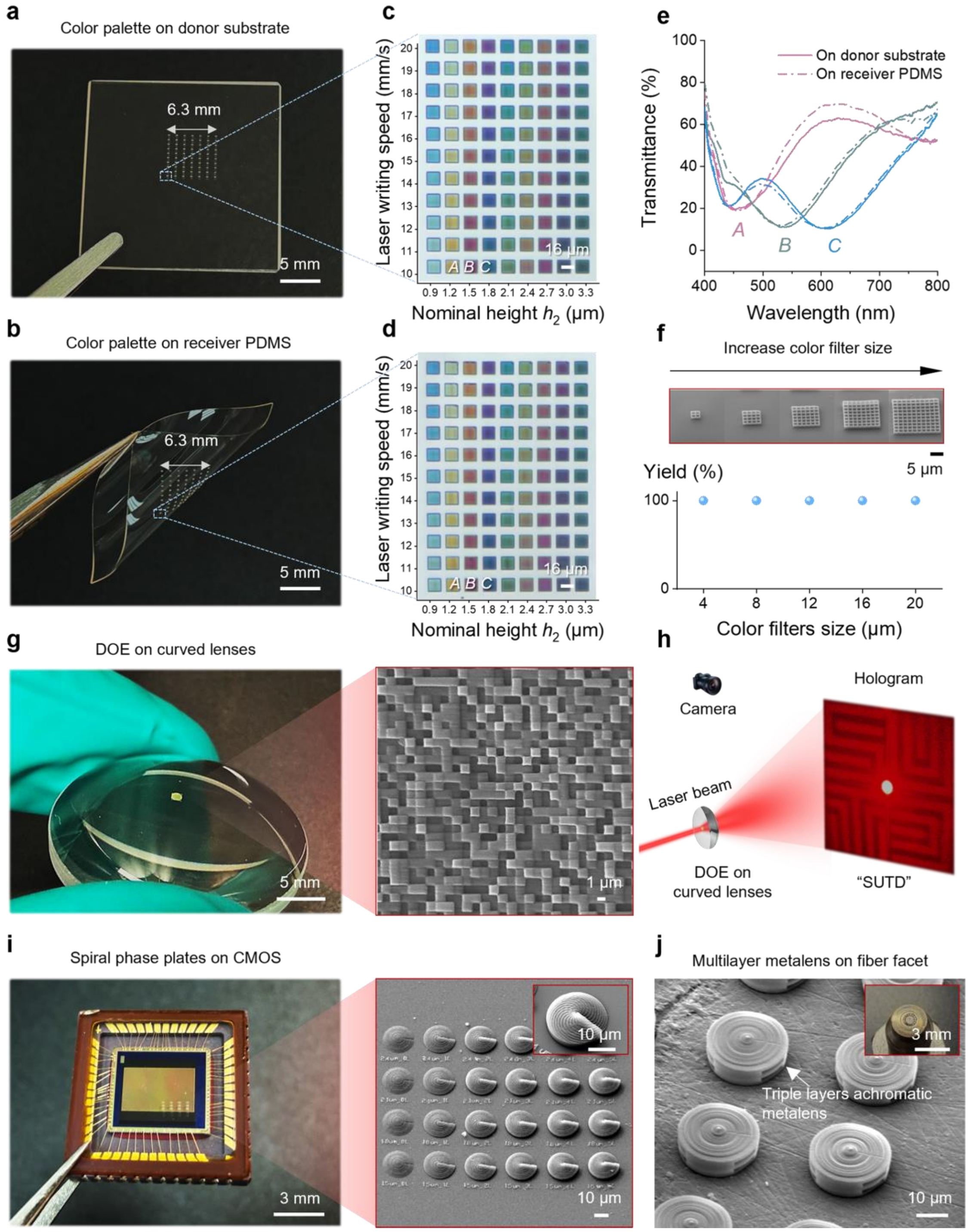

**Fig. 3. Transfer printing of 2.5D photonic crystals and optical elements.** (a-b) Large-area color palette array (6.3 mm × 5.6 mm) with 6,930 elements transferred from quartz onto PDMS, filter size: 16 μm. (c-d) Optical images of 2.5D color palette unit before and after transfer printing. (e) Representative transmittance spectra labeled *A*, *B*, and *C*. (f) Transfer printing yield of color filters for each size ranging from 4 to 20 μm. (g) Demonstration of integrating diffractive optical elements (DOEs) with curved lenses and SEM image of phase pixels of the DOE. (h) The DOE illuminated by a laser beam generates a "SUTD" holographic pattern. (i) Demonstration of integrating spiral phase plates arrays on complementary metal-oxide-semiconductor (CMOS). (j) SEM of multilayer achromatic metalens arrays on plastic optical fiber facet, the inset figure shows macroscopic

end-face structure of a multi-bundle plastic optical fiber. The structures shown in Fig. 3 were fabricated using IP-DIP2 resist.

Our process provides a route to fabricating uniform inorganic nanostructures, which is particularly suitable for high-resolution 3D inorganic photonic crystals (e.g. glass-based structures). While 3D photonic crystals fabricated by TPL have been widely applied in nanophotonics and related fields, achieving perfectly ordered inorganic 3D photonic crystals with high uniformity and structural fidelity remains complicated. Typically, hybrid resists were patterned via TPL and then shrunk at high temperature to remove organic components[21], but the resulting sub-diffraction non-uniform inorganic architectures such as visible-range photonic crystals, generally exhibit desired optical properties by sacrificing underlying structures[4]. While sacrificial-layer approaches could potentially fabricate fully uniform inorganic structures, they introduce additional fabrication complexity and often compromise yield and structural stability. Conversely, the uniform print-and-shrink process enables the realization of defect-free, perfectly ordered inorganic 3D photonic crystals without using complex wet-processing steps (Fig. 4a). As shown in Fig. 4b, we fabricated diamond-lattice photonic crystals with an initial lattice spacing of ~1.53 μm using a $SiO_2$ resist (Glass-Nano resin[4]). Shrinking at 550 °C induced isotropic shrinkage by a factor of ~4.4, yielding a final spacing of ~347 nm and excellent 3D uniformity (Fig. 4c). Measurements at 20 randomly selected surface points verified the excellent uniformity of the thermally shrunk lattices (Supplementary Fig. 18). By adjusting the initial lattice parameters, the structural colors of the 3D photonic crystals can be systematically tuned, as confirmed by reflectance spectra (Fig. 4d and Supplementary Fig. 18).

Extending from 3D photonic crystals, the manipulation of photons at the mesoscopic scale forms the foundation of photonic integrated circuits. A variety of functional photonic microstructures, including basic waveguides and cavities as well as more complex coupled architectures, have been developed to confine, guide, and manipulate photons[42]. Among them, whispering-gallery-mode (WGM) resonators, capable of storing light within an ultracompact volume for extended durations, enable strong field confinement, enhanced light-matter interactions, and substantial intracavity power buildup (Fig. 4e)[43]. As the resonator diameter decreases, the WGM wavelengths shift into the visible and ultraviolet regimes, imposing significant challenges on fabrication of 3D nanoscale fully dielectric structures. A feasible fabrication pathway is enabled by our process, through which WGM resonator arrays with diverse geometries are realized, including ~2 μm-diameter microgoblets, microdisks, and vertical-slit-engineered microdisks, all featuring nanoscale cavity thicknesses and defect dimensions (Fig. 4f and Supplementary Fig. 19). Numerical analysis of the microgoblet confirms the formation of visible-wavelength resonance modes with circulating fields that maintain a curved surface-parallel polarization (Fig. 4g).

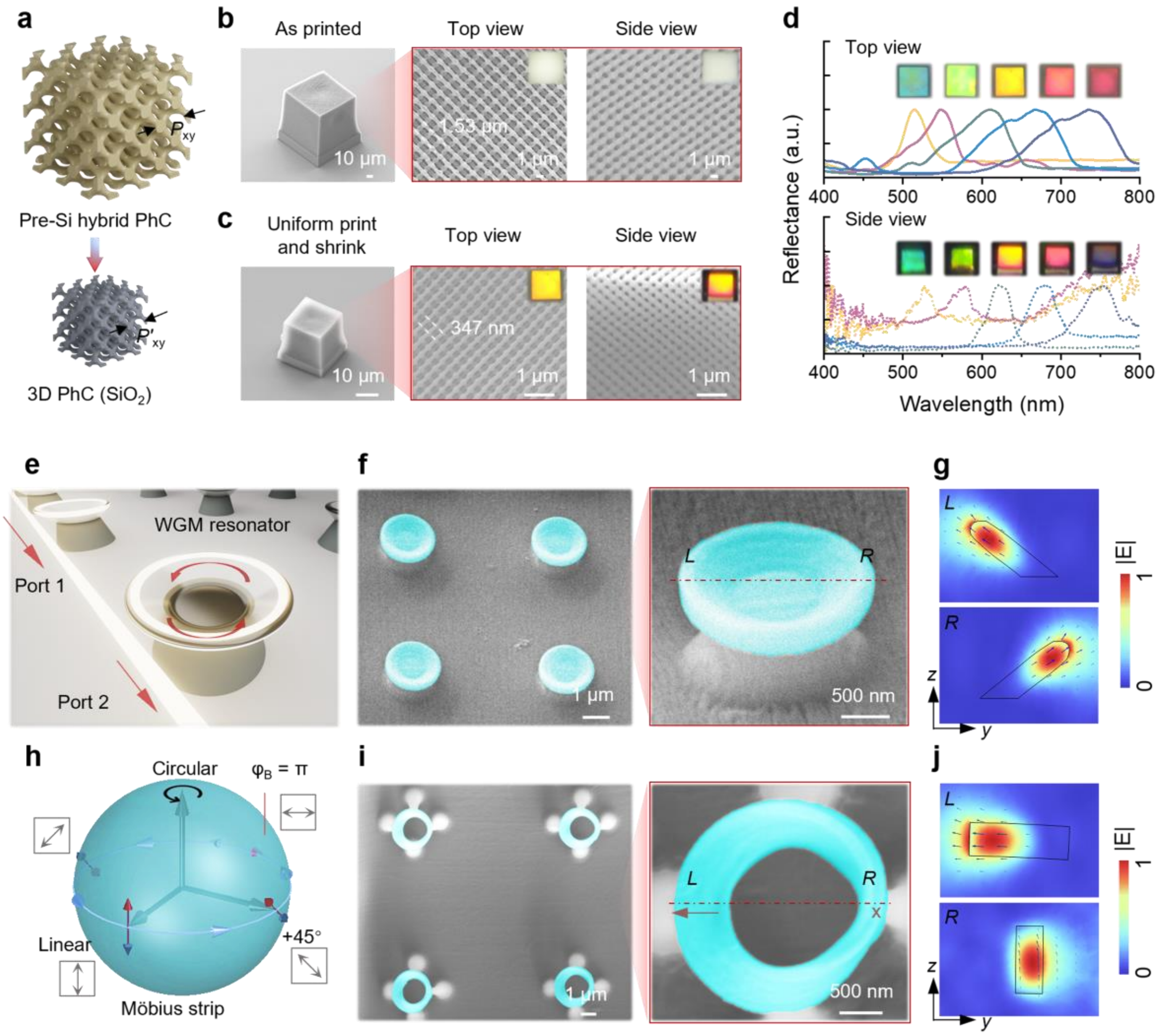

**Fig. 4. Glass 3D photonic crystals and optical resonators for photon manipulation.** (a) Schematic of 3D photonic crystals before and after uniform shrinkage, with the lateral pitch ($P_{xy}$) shifted into the visible wavelength range. (b) SEM images of as-printed diamond-lattice photonic crystals and (c) after uniform shrinkage (top and side views); optical micrographs of photonic crystals (inset). (d) Reflectance spectra and optical micrographs of 3D glass photonic crystals with varying lattice constants. (e) Illustration of the whispering-gallery-mode (WGM) resonators. (f) The fabricated microgoblet WGM resonator array by $SiO_2$. (g) Simulated electric field distribution in microgoblet WGM resonator. (h) Topological structures and polarization evolution of the mode of light confined in a Möbius strip on the Poincaré sphere. (i) The fabricated Möbius strip resonator array by $SiO_2$. (j) Simulated electric field distribution in Möbius strip resonator.

Geometric topology has also emerged as a powerful framework for advancing photon manipulation. The Möbius strip, characterized by its one-sided topology, introduces a geometric (Berry) phase through spin–orbit coupling when light propagates along its twist, leading to non-trivial mode evolution and the theoretical emergence of half-integer resonances between conventional whispering-gallery modes (Fig. 4h)[44,45]. Realization of an ideal Möbius strip requires a width far below the considered light wavelength in the waveguiding medium to ensure strict field confinement, which is challenging with current fabrication techniques. Exploiting 3D inorganic structures capability of our process, a ~2 μm-diameter Möbius resonator array with a 180 nm strip thickness is fabricated (Fig. 4i). Simulations confirm that polarization rotation of the waveguided light along the Möbius path, from parallel (left) to perpendicular (right) (Fig. 4j). Thus, our strategy potentially establishes a 3D nanofabrication route for topological photonic structures and enables new possibilities for topological information processing and coherent wave dynamics in quantum photonics.

**Discussion**

This sublime transfer printing strategy for manipulation and construction of high-resolution 3D nanostructure, enables conformal, damage-free transfer of ensembles of arbitrary 3D structures onto a whole range of diverse substrate types. We revealed that complete encapsulation of 3D structures in naphthalene, followed by its sublimation into the gas phase, eliminates stress concentration and prevents damage from capillary forces. Theoretical analysis and mechanical tests reveal the underlying adhesion-control mechanism for the damage-free process, where low surface energy yields near-zero interfacial adhesion, enabling transfer and uniform shrinkage, while stress is negligible and adhesion competition dominates during sublimation printing. This approach enables high-yield transfer of 3D structures onto arbitrary substrates, including flexible, suspended, and low-surface-energy surfaces, without chemical treatment. It further enables the uniform print-and-shrink process for the fabrication of inorganic 3D architectures with sub-100 nm feature size (e.g. 80 nm line width). We demonstrated various nanostructures and functional photonic devices including 2.5D/3D photonic crystals, curved DOEs, on-chip spiral phase plates, multilayer achromatic fiber-metalens, WGM resonator and Möbius strip topological resonator. Subsequent studies may explore strategies to achieve conformal release over geometrically complex and irregular surfaces. While there are some key challenges such as precision alignment in this area, we believe our approach here enables the application of transfer printing in various fields that require submicron scale resolution in three dimensions. Sublime transfer print could find use in adjacent fields such as 2D materials, 3D integration, bioelectronics, 3D topological photonics and quantum devices.

**Acknowledgements**

J.K.W.Y acknowledges National Research Foundation (NRF) Singapore (NRF-NRFI06-2020-0005). H.D. acknowledges National Natural Science Foundation of China (52425508). This research is also partially supported by grants from the National Research Foundation, Prime Minister's Office, Singapore under its Campus of Research Excellence and Technological Enterprise (CREATE) programme. H.W acknowledges the support from National Natural Science Foundation of China (606HWRC2025117001), the Fundamental Research Funds for the Central Universities (501RCQD2025117002), and the Research Funding of Hangzhou International Innovation Institute, Beihang University (015733207-000001).

**Author Contributions**

L.C., H.D. and J.K.W.Y. conceived the idea. L.C., H.W. and W.Z. designed the experiments, fabricated and characterized the samples. F.F., P.L. and X.B. contributed theoretical analysis and mechanical tests. J.Y.E.C., R.Y.T., H.T.W., C.-F.P., B.W., and Z.L. assisted in the characterization and data analysis. J.K.W.Y. supervised the research. All authors contributed to the writing and revision of the manuscript. L.C. and H.W. contributed equally to this work.

**Competing interests**

The authors declare no competing interests.

**Additional information**

Supplementary Information is available for this paper.

**Correspondence** and requests for materials should be addressed to J.K.W.Y.